\documentclass[useAMS,usenatbib]{mn2e}

\usepackage{graphicx}
\usepackage{amssymb}
\usepackage{natbib}

\def\boldd{\mathrm{d}}           
\def\i{\rm i}                
\usepackage{epstopdf}
\usepackage{pgfplots}
\usepackage{bm}
\usepackage{amsmath}
\usepackage{rotating}

\usepackage{color}
\long\def\chng#1{{#1}}       

\title[The thermal emission from boulders]{The thermal emission from boulders on (25143) Itokawa and general implications for the YORP effect}
\author[P. \v Seve\v cek et~al.]{P. \v Seve\v cek$^{1}$\thanks{E-mail: sevecek@sirrah.troja.mff.cuni.cz},
 M. Bro\v z$^{1}$,
 D. \v Capek$^2$,
 J. \v Durech$^1$
 \\
$^{1}$Institute of Astronomy, Charles University, Prague, V Hole\v sovi\v ck\'ach 2, 18000 Prague 8, Czech Republic\\
$^{2}$Astronomical Institute of the Academy of Sciences, Fri\v cova 298, 251 65 Ond\v rejov, Czech Republic}

\begin{document}

\date{Accepted 2015 March 31. Received 2015 March 30.; in original form 2014 September 21}

\pagerange{\pageref{firstpage}--\pageref{lastpage}} \pubyear{2015}

\maketitle

\label{firstpage}


\begin{abstract}
Infrared radiation emitted from an asteroid surface causes a torque that can
significantly affect rotational state of the asteroid. The influence of small
topographic features on this phenomenon, called the YORP effect, seems to be 
of utmost importance. In this work, we show that a lateral heat diffusion
in boulders of suitable sizes leads to an emergence of a~local YORP
effect which magnitude is comparable to the YORP effect due to the global
shape. We solve a~three-dimensional heat diffusion equation in a~boulder and
its surroundings by the finite element method, using the FreeFem++ code. The
contribution to the total torque is inferred from the computed temperature
distribution. Our general approach allows us to compute the torque induced 
by a~realistic irregular boulder.
For an idealized boulder, our result is consistent with an existing
one-dimensional model. We also estimated (and extrapolated) a~size
distribution of boulders on (25143)~Itokawa from close-up images of its surface.
We realized that topographic features on Itokawa can potentially induce a~torque corresponding to
a~rotational acceleration of the order $10^{-7}\,{\rm rad}\,{\rm day}^{-2}$ and can therefore
explain the observed phase shift in light curves.
\end{abstract}

\begin{keywords}
    minor planets, asteroid: individual: (25143) Itokawa -- methods: numerical.
\end{keywords}



\section{Introduction}\label{sec:introduction}

The Yarkovsky-O'Keefe-Radzievskii-Paddack effect is the torque 
caused by the infrared emission from an asteroidal surface 
that has an influence on a~rotational state of an asteroid \citep{Rubincam_2000}. 
It is now widely recognized as an important factor, 
affecting the evolution of rotational states of asteroids 
alongside mutual collisions and tidal torques. 
The YORP effect helped to explain numerous observed phenomena, 
such as a~spin axis alignment of asteroids in the Koronis family \citep{Vokrouhlicky_2003}, 
a~non-maxwellian rotational frequency distributions of small main-belt asteroids \citep{Pravec_2008} 
or significant binary asteroid population among near-Earth objects \citep{Walsh_2012}.

Even a~\emph{direct} evidence of a~non-gravitational torque has been found. 
A~phase shift in light curves has been measured for a~few asteroids 
that cannot be explained by a~solely gravitational model --- %
(1862) Apollo \citep{Kaasalainen_2007}, 
(54509) YORP \citep{Lowry_2007, Taylor_2007}, 
(1620) Geographos \citep{Durech_2008a}, 
(3103) Eger \citep{Durech_2012}, 
and finally, (25143)~Itokawa \citep{Lowry_2014}.
There is no known mechanism except the YORP effect that 
    could explain the quadratic trend in 
    rotational phase observed for these asteroids.

The asteroid Itokawa has been a~suitable candidate for a~detection of YORP effect 
for its highly asymmetric shape and its favourable position among near-Earth objects. 
\cite{Vokrouhlicky_2004} predicted a~measurable acceleration of rotation 
of the order of $10^{-7} \,\rm rad\,day^{-2}$, 
based on the shape model derived by radar ranging. 
Itokawa was then a~target of the Hayabusa spacecraft in 2005 and 
a~state-of-the-art shape model of the asteroid was constructed 
from silhouette images \citep{Gaskell_2006}. 
The torque computed using the latter model would lead to a~significant deceleration 
$-(1.8 {\rm \ to\ }3.3)\times 10^{-7}\,\rm rad\,day^{-2}$ \citep{Scheeres_2007}.
However, the measured phase shift in light curves revealed an \emph{acceleration}  
$+(0.35\pm 0.04)\times 10^{-7} \,\rm rad\,day^{-2}$ \citep{Lowry_2014}. 
Theoretical models did not predict even the sign of the effect correctly. 
This discrepancy between the observed and predicted change of 
    the angular frequency is concerning and only one viable explanation 
    has been put forward to date.%

The observed rotational acceleration could be attributed 
to density inhomogeneities in the asteroid. 
\cite{Scheeres_2008} showed that the YORP effect on Itokawa is 
indeed sensitive to the position of the center of mass. 
Based on the measured acceleration, \cite{Lowry_2014} computed 
the required offset between the center of mass and the center of figure 
to be ${\sim}\,21\,\rm m$. 
Such offset indicates that the asteroid might consist of 
two parts with substantially different densities --- %
$(2850\pm500) \,\rm kg\,m^{-3}$ and $(1750\pm110) \,\rm kg\,m^{-3}$.

The deceleration predicted by \cite{Scheeres_2007} was computed 
from the shape model with $\sim 50000$ facets. 
Calculations of the effect with a~more detailed shape 
lead to an even bigger deceleration.
According to \cite{Breiter_2009}, the deceleration does not show 
any sign of convergence 
with increasing resolution, implying that even sub-meter sized 
surface features possibly have a~non-negligible influence.
For the highest available shape resolution, their model 
    predicted the deceleration $-5.5 \times 10^{-7}\,\rm rad \,day^{-2}$.%

\cite{Lowry_2014} employed the advanced thermophysical model 
    \citep{Rozitis_2011, Rozitis_2012, Rozitis_2013}, including 
    the effect of thermal infrared beaming 
    and the global self-heating of the asteroid. 
By varying the distribution of rough surface in patchy ways, 
the model showed a change of angular velocity $(-1.80 \pm 1.96) \times 10^{-7}\,\rm rad \, day^{-2}$
(see Fig.~5 in the cited paper). 
Remarkably, an acceleration can be achieved even without 
a shift of the center of mass.
However, this result was obtained only in $16.5\%$ cases and 
the roughness distribution corresponding to these cases seems rather artificial.%

There is a~problem that shapes with surface features of sub-meter sizes 
cannot be easily included in existing models of the YORP effect. 
There are several reasons for this limitation. 
First, numerical YORP models typically assume that temperature changes 
only in the direction perpendicular to the surface 
(i.e. a~plane-parallel approximation). 
This assumption allows a~solution of the one-dimensional heat diffusion equation 
for each surface facet independently. 
It is well justified as long as surface features are significantly
larger then the diurnal thermal skin depth, 
which varies from mm to dm \citep{Vokrouhlicky_1999}. 
This assumption is no longer applicable for a~high-resolution shape model 
and a~full three-dimensional solution of the heat diffusion equation is required.
Second, no shape is described to the required level of detail. 
So far, the best shape model is that of the asteroid Itokawa. 
The model in the best available resolution consists of over 3 million facets, 
which corresponds to meter-sized surface features \citep{Gaskell_2006}. 

As \cite{Golubov_2012} pointed out, surface features of sizes comparable to 
the thermal skin depth could potentially have 
significant influence on the total YORP effect. 
They considered a~stone wall (an idealized boulder) located on the equator of 
a~spherical asteroid and aligned with a~local meridian. 
The wall was assumed to be high enough so that the heat would be 
mostly conducted in a~transverse direction and the heat diffusion equation 
can be solved using the one-dimensional approximation. 
They demonstrated that the emission from the surface of the wall can 
create a~torque that will not vanish after averaging over the rotational period. 
Assuming a~large number of such ``walls'' placed along the equator, 
the corresponding torque may be comparable to the torque arising from the global-shape asymmetry.

\cite{Golubov_2014} generalized the previous model by assuming spherical boulders. 
    They studied a dependence of the torque on a number of free parameters of the problem.
    As expected, the resulting torque is lower than in a simple one-dimensional model;
    nevertheless, boulders can still affect rotational dynamics significantly.
    Authors assumed even mutual shadowing and heating of the boulders.%

The goal of this paper is to solve the heat diffusion equation in a~\emph{realistic} boulder. 
The problem requires a~numerical solution in a~general three-dimensional domain. 
We derive a~formulation of the numerical problem in Section~\ref{sec:HDE}. 
We discuss the magnitude of the torque induced by a~single boulder in Section~\ref{sec:Pi}. 
We estimate the total torque that boulders contribute to the YORP effect on the asteroid Itokawa in Section~\ref{sec:itokawa}. 
Finally, the results of our model and the implications are summarized in Section~\ref{sec:conclusions}.


\section{The heat diffusion equation and a~weak formulation of the problem}\label{sec:HDE}
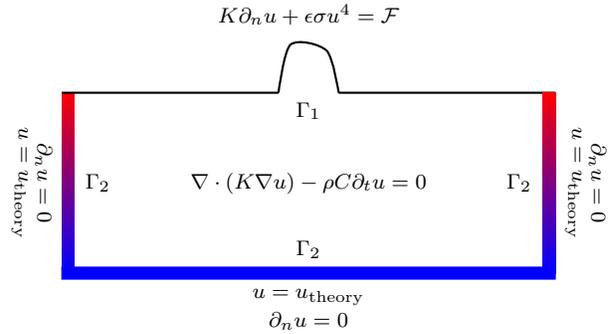
\begin{figure}
    \centering
    \begin{tikzpicture}[scale=0.8]
       \filldraw[color=blue] (4.1cm, -2.9cm) rectangle (-4.1cm, -3.1cm);
       \shade[shading=axis,bottom color=blue,top color=red] (-4.1cm, -3cm) rectangle (-3.9cm, 0cm); 
  
     \draw [thick] (-4.1cm, 0cm) -- (-0.5cm, 0cm);
     \draw [thick] plot [smooth] 
        coordinates { (-0.5cm, 0cm)  (-0.3cm, 0.8cm)  (0.3cm, 0.7cm)  (0.5cm, 0cm) };

     \draw [thick] (0.5cm, 0cm) -- (4.1cm, 0cm);
     \shade[shading=axis,bottom color=blue,top color=red] (4.1cm, -3cm) rectangle (3.9cm, 0cm); 

     \node [] at (0cm, 1.3cm) { $K \partial_n u+ \epsilon \sigma u^4 = \mathcal{F} $ };
     \node [rotate=-90] at (-4.8cm, -1.5cm) { $u=u_{\rm theory}$ };
     \node [rotate=-90] at (-4.4cm, -1.5cm) { $\partial_n  u= 0$ };
     \node [rotate=-90] at (4.4cm, -1.5cm) { $u=u_{\rm theory}$ };
     \node [rotate=-90] at (4.8cm, -1.5cm) { $\partial_n  u= 0$ };
     \node [] at (0cm, -3.4cm) { $u=u_{\rm theory}$ };
     \node [] at (0, -3.8cm) { $\partial_n u= 0$ };
     \node [] at (0, -0.3cm) { $\Gamma_1$ };
     \node [] at (0, -2.6cm) { $\Gamma_2$ };
     \node [] at (3.5cm, -1.5cm) { $\Gamma_2$ };
     \node [] at (-3.5cm, -1.5cm) { $\Gamma_2$ };

 \node [] at (0,-1.5cm) { $ \nabla \cdot (K \nabla u) - \rho C {\partial_t u} = 0$ };
    \end{tikzpicture}
    \caption{The domain $\Omega$ and boundary conditions of our problem. 
        The boulder is located on the top.
        The strips indicate the boundary $\Gamma_2$, 
        where the temperature is kept by a~Dirichlet boundary condition. 
        The temperature distribution inside the domain and on the surface~$\Gamma_1$ 
        is then computed numerically as a~solution of Eq. (\ref{HDE}) with boundary conditions, 
        Eqs.~(\ref{BC}) to~(\ref{E_sc}). } 
    \label{img:domain}
    \end{figure}

    \begin{table}
    \centering
    \begin{tabular}{ll}
        \hline
        Notation & Meaning \\
        \hline
        $u$ & temperature \\
        $\hat u$ & numerical solution \\
        $\mathcal{L}$ & differential operator corresponding to \\
                      & the heat diffusion equation \\
        $N_j$ & basis functions \\
        $W_j$ & weighting functions \\
        $u_j$ & coefficients of the decomposition of $\hat u$\\
        $K$ & thermal conductivity \\
        $\rho$ & density \\
        $C$ & specific heat capacity \\
        $\epsilon$ & infrared emissivity \\
        $A$ & hemispherical albedo \\
        $\sigma$ & Stefan--Boltzmann constant \\
        $\omega$ & angular frequency \\
        $P$ & rotational period \\
        $c$ & speed of light \\
        $\Phi$ & solar flux at the distance of the asteroid\\
        $\mathcal{F}$ & flux absorbed by the surface \\
        $L$ & diurnal thermal skin depth \\
        $\Theta$  & thermal parameter \\
        $u_\star$ & subsolar temperature \\
        $\Pi$ & dimensionless pressure \\
        $\mathcal{V}$ & latitudinal dependence of $\langle \Pi\rangle$ \\
        $\Omega$ & computational domain \\
        $\Gamma_1$, $\Gamma_2$ & boundaries of the domain \\
        $\vec r$ & position vector \\
        $t$ & time \\
        $\vec n$ & outward normal vector \\
        $\vec s$ & body--Sun direction \\
        $\partial_n$ & derivative along the (outward) normal \\
        $\mu$ & shadowing function \\
        $\nu$ & visibility function \\
        $H$ & Heaviside step function \\
        $\vec f$ & force \\
        $\vec \tau$ & torque \\
        $\vec e$ & rotational axis \\
        $\vartheta$ & asteroidal latitude \\
        $\ell$ & characteristic size of the boulder \\
        $I$ & moment of inertia of Itokawa \\
        $N(\ell)$ & differential size distribution of boulders\\
        $\gamma$ & exponent of the size distribution\\
        \hline
        \end{tabular}
        \caption{Notation used in the paper.}
        \label{tab:notation}
    \end{table}

Our problem may be specified as follows.
We search for a~temperature $u(\vec r, t)$ inside the boulder and its
surroundings, i.e. an unknown scalar function on a~domain~$\Omega$. The
differential operator corresponding to the heat diffusion equation (HDE) is:
\begin{equation}
{\cal L} \equiv \rho C \partial_t - \nabla \cdot K \nabla\,,\label{L}
\end{equation}
where $K$~denotes the thermal conductivity,
$\rho$~the density,
$C$~the specific heat capacity of the material.
The function~$u$ thus has to fulfill the relation:
\begin{equation}
{\cal L}(u) = 0\,.\label{HDE}
\end{equation}
At the same time, we require the boundary conditions to be met at the boundary $\partial \Omega$ of the domain. 

\subsection{Boundary conditions}
The boundary consists of two parts denoted $\Gamma_1$ and $\Gamma_2$, as shown in Figure~\ref{img:domain}.
The boundary $\Gamma_1$ represents the surface of the asteroid, the boundary condition is essentially an energy balance equation:
\begin{equation}
    K \partial_n u + \epsilon\sigma u^4 = \mathcal{F} \,, 
    \label{BC}
\end{equation}
where $\partial_n$~denotes a~derivative along the normal,
$\epsilon$~the infrared emissivity,
$\sigma$~the Stefan--Boltzmann constant,
$\mathcal{F}$ the incoming flux absorbed by the surface.
The boundary $\Gamma_1$ is a~non-convex surface, 
thus the radiative heat exchange also contributes to the total flux $\mathcal{F}$ 
(i.e. the self-heating effect). 
We denote the solar flux contribution $\mathcal{F}_\odot$ and 
the contribution of flux coming from visible parts of the surface --- %
either the thermally emitted flux or the scattered flux --- %
as $\mathcal{F}_{\rm th}$ and $\mathcal{F}_{\rm sc}$ respectively. 
The total absorbed flux is the sum
$\mathcal{F} = \mathcal{F}_\odot + \mathcal{F}_{\rm th} + \mathcal{F}_{\rm sc}$. 
These terms can be expressed as:
\begin{align}
    \mathcal{F}_\odot &= (1-A)\Phi \mu \vec s \cdot \vec n \,, \\
 \label{E_rad} \mathcal{F}_{\rm th}& = (1-A) \int_{\Gamma_1}\!\! \epsilon' \sigma u'^4 {\cos \alpha \cos \alpha' \over \pi (\vec r-\vec r')^2}\nu \, \boldd \Gamma' \,, \\
  \label{E_sc} \mathcal{F}_{\rm sc}& = (1-A) \int_{\Gamma_1} \!\! A' \Phi \mu' \vec s \cdot \vec n'  {\cos \alpha \cos \alpha' \over \pi (\vec r-\vec r')^2} \nu \, \boldd \Gamma' \,, 
\end{align}
where 
$A$~is the hemispherical albedo,
$\Phi$~the flux of solar radiation,
$\vec s$~the body--Sun direction,
$\vec n$~the outward normal to the surface,
$\vec r$~the position vector,
$\alpha$~the angle between the local normal and the direction vector connecting points $\vec r$ and $\vec r'$, 
$\mu$ the shadowing function,
$\nu$ the visibility function (see Table \ref{tab:notation}).
The prime denotes a~value of a~quantity at the point of surface element $\boldd \Gamma'$. 
We assume the Lambert's cosine law for the intensity of thermal emission and scattered radiation, 
hence the $\cos \alpha'$ in equations (\ref{E_rad}) and (\ref{E_sc}). 

The shadowing function $\mu$ is defined on the surface $\Gamma_1$. 
The value of $\mu(\vec r)$ equals 1 if the point $\vec r$ is insolated, 
0 if it lies in the shadow. 
The visibility function $\nu$ is defined on $\Gamma_1 \times \Gamma_1$ space. 
We assign the value of function $\nu(\vec r, \vec r')$ to 1 if points $\vec r$ and $\vec r'$ have a~visual contact, 
0 otherwise. In most cases, the value of $\nu$ is simply  
$\nu(\vec r, \vec r') = H(\vec n \cdot (\vec r - \vec r')) H(\vec n' \cdot (\vec r' - \vec r) )$,
where $H$ is the Heaviside step function and $\vec n$, $\vec n'$ denote the local normal at points $\vec r$ and $\vec r'$.

The discretized boundary $\Gamma_1$ is defined by a~set $S_i$ of triangular facets. Integrals in equations (\ref{E_rad}) and (\ref{E_sc}) can be therefore computed as sums, $\int_{\Gamma_1} \boldd \Gamma' \rightarrow \sum_i S_i$. The values of the shadowing function $\mu$ and the visibility function $\nu$ always correspond to whole facets.  This restriction gives rise to an error; however, it can be estimated and limited substantially by choosing a~high-resolution surface mesh.

We also need to specify conditions on the boundary $\Gamma_2$, which goes through the interior of the asteroid, closing the boundary $\partial \Omega$. It can be selected arbitrarily; we choose a~shape corresponding to five walls of a~block, which is a~convenient choice as we can simply set a~zero-flux boundary condition: 
\begin{equation}
    K \partial_n u = 0  \,.
    \label{BC2}
\end{equation}
This condition will be met as long as dimensions of the domain $\Omega$ are significantly greater than dimensions of the boulder. The influence of the boulder can be considered negligible at large distances. At sides of the domain, the temperature will only change in the direction perpendicular to the surface, the dot product of the normal vector and the temperature gradient will therefore be null. At great depths, the temperature will be effectively constant, which means the temperature gradient at
the bottom of the domain will be null, satisfying the boundary condition (\ref{BC2}).

We also need to specify an initial condition as the HDE is an evolution equation. However, we seek for a~stationary solution that does not depend on a~chosen initial condition. The choice of a~condition will affect the speed of convergence only. The best available estimate of the solution is the linearized analytical solution in a~half-space domain, derived in Appendix~\ref{sec:analytic}, hence we set:
\begin{equation}
u = u_{\rm theory}, \ t=0 \,.
\end{equation}


\subsection{A discretization}

We are going to solve the HDE (Eq. \ref{HDE}) numerically, using a~finite-element discretization in space.
In this approach, the function~$u$ is approximated by \citep{Langtangen_2003}:
\begin{equation}
u \doteq \hat u = \sum_{j=1}^M u_j N_j\,,\label{hatu}
\end{equation}
where $N_j$~denote prescribed basis functions,
$u_j$~unknown coefficients we search for
and $M$~corresponds to the number of vertices defined on the domain.
Since $\hat u$ is only an approximation of $u$,
applying the operator would generally yield a~non-zero result:
\begin{equation}
{\cal L}(\hat u) \ne 0\,,
\end{equation}
nevertheless, we require the integral of all residua
over the domain to be zero:
\begin{equation}
\int_\Omega {\cal L}(\hat u)W_i{\rm d}\Omega = 0\,,
\end{equation}
where $W_i$ are suitable weighting (test) functions.
This is called a~{\em weak formulation\/} of the problem.
In the Galerkin method, the test functions are simply the basis functions,
$W_i \equiv N_i$, so that:
\begin{equation}
\int_\Omega {\cal L}(\hat u)N_i{\rm d}\Omega = 0\,.
\end{equation}
Essentially, this constitutes a~system of~$M$ equations for $u_j$ coefficients.

In our case of the HDE~(Eq.~\ref{HDE}):
\begin{equation}
\int_\Omega \rho C \partial_t \hat u N_i {\rm d}\Omega - \int_\Omega \nabla\cdot (K\nabla\hat u) N_i {\rm d}\Omega = 0\,.
\end{equation}
The second term may be rewritten according to the Green lemma as:
\begin{equation}
    \int_\Omega \nabla\cdot (K\nabla\hat u) N_i {\rm d}\Omega = -\!\!\int_\Omega\! K\nabla\hat u\cdot\nabla N_i{\rm d}\Omega\, + \oint_{\partial \Omega}\!\! K\partial_n\hat u N_i {\rm d}\Gamma,\,\,
\end{equation}
which enables to incorporate the boundary condition easily,
because we can express the normal derivative from boundary conditions (\ref{BC}) and (\ref{BC2}), that is 
$K\partial_n \hat u = -\epsilon\sigma\hat u^4 + \mathcal{F}$ on $\Gamma_1$,
$K\partial_n \hat u = 0$ on $\Gamma_2$.

Regarding the temporal derivative, we use a~finite-difference discretization:
\begin{equation}
\partial_t \hat u \simeq {\hat u^n - \hat u^{n-1}\over\Delta t}
\end{equation}
and an implicit Euler scheme, so that we plug $\hat u^n$ in the remaining terms,
whenever possible. The only exception is the non-linear radiative term,
where we perform a~linearization:%
\begin{equation}
    (\hat u^{n,m})^4 \simeq (\hat u^{n,m-1})^3 \hat u^{n,m} 
\end{equation}%
and we employ an iterative method to find a~solution;
at the given time-step (denoted by superscript $n$) 
    we find a sequence of solutions $\hat u^{n,m}$ to a linear problem,
    until $|\hat u^{n,m}-\hat u^{n,m-1}|$ is sufficiently small.
    The initial value $\hat u^{n,0}$ can be selected arbitrarily, 
    a common choice is $\hat u^{n,0} = \hat u^{n-1}$.

In some cases, the iterative method does not converge and 
we thus introduce a~relaxation parameter $\zeta$. 
In each iteration, we find a~preliminary solution $\hat u^{n,m}_*$ of 
the linear problem and we then assign a~new value of $\hat u^{n,m}$ 
by taking a~linear combination of the current and previous solutions:
\begin{equation}
    \hat u^{n,m} \equiv \zeta \hat u_*^{n,m} + (1-\zeta) \hat u^{n,m-1} \,.
    \label{relaxation}
\end{equation}%
We achieved a~convergence for all considered values of parameters by selecting $\zeta =0.6$.

The final equation is thus:
$$\int_\Omega {\rho C\over\Delta t} \hat u^{n,m} N_i\, {\rm d}\Omega
- \int_\Omega {\rho C\over\Delta t} \hat u^{n-1} N_i \,{\rm d}\Omega
+ \int_\Omega\! K\nabla\hat u^{n,m}\cdot\nabla N_i\,{\rm d}\Omega \,+$$
\begin{equation}
    + \int_{\Gamma_1} \!\!\epsilon\sigma (u^{n,m-1})^3 \hat u^{n,m} N_i \,{\rm d}\Gamma
- \int_{\Gamma_1} \!\! \mathcal{F}\,\boldd\Gamma = 0\,.
\label{weak_problem}
\end{equation}%
We actually need not to substitute for~$\hat u$ from Eq.~(\ref{hatu})
or express the corresponding matrices, because this is done automatically
by the FreeFem++ code \citep{FreeFem}.
We use a~conjugate gradient method for the matrix inversion,
which is suitable for sparse linear systems.

After a careful testing of our numerical method (see Appendix~\ref{sec:uncertainty}),
we choose the time step $\Delta t = 10^{-3} P$,
where $P$ is the period present in the insolation function~${\cal F}$.
The spatial step is controlled by the maximum volume $\Delta\Omega = 10^{-5}\,{\rm m}^3$
of tetrahedra generated by the TetGen code \citep{TetGen}.


\section{The mean torque caused by an irregular boulder}\label{sec:Pi}

The magnitude of a~recoil force varies during a~rotational period and a~revolution of an asteroid around the Sun. 
A~long-term effect of the force is therefore given by its time-averaged value. 
We follow the assumption of \cite{Golubov_2012} and consider an asteroid on 
a~circular orbit with zero obliquity. 
\chng{In fact, the eccentricity of Itokawa is $e=0.28$; 
we address this issue in Section~\ref{sec:eccentricity}.}
Although the YORP effect depends on 
the obliquity in a~non-trivial way \citep{Capek_2004}, the zero obliquity 
allows us to average the recoil force over a~rotational period only. 

We consider the thermal emission and the scattered radiation. The direct radiation pressure has a~negligible influence \citep{Nesvorny_2008}. Again, we assume the Lambert's cosine law for the intensity of scattered and emitted radiation. 
The recoil force from the surface element $\boldd S$ is then:
\begin{align}
    \label{f_rad}
    \boldd \vec f_{\rm th} &= -{2 \over 3}{\epsilon \sigma \over c} u^4 \vec{n}\, \boldd S \,,\\
    \boldd \vec f_{\rm sc} &= -{2 \over 3} {A \Phi \over c} \mu (\vec s \cdot \vec n ) \vec n\, \boldd S \,.
\end{align}
The total torque caused by the boulder is given by the surface integral over the boulder:
\begin{equation}
\vec \tau = \int_{\Gamma_1} \!\! \vec r \times \boldd \vec f\,. 
\end{equation}

The direction of the torque is generally different from the axis of rotation $\vec e$. Both the direction and the magnitude of the torque depend on exact shape of the boulder. However, even a~symmetric boulder can induce a~non-zero torque due to the lateral heat diffusion. The torque is caused by the asymmetry of emission from the eastern side and western side of the boulder. 
We anticipate the torque will therefore have a~direction of the rotational axis $\vec e$.


\subsection{The coordinate system and free parameters of the problem}
\label{sec:coordinates}

We choose a~topocentric coordinate system centered on the studied boulder. The $z$ axis has therefore a~direction of a~local normal, $x$ axis is aligned with a~meridian and $y$ axis completes a~right-handed orthogonal Cartesian system. 

We introduce quantities that help us to reduce a~number of independent parameters of the problem. We define the subsolar temperature:
\begin{equation}
    u_\star \equiv \sqrt[4]{(1-A)\Phi \over \epsilon\sigma} \,,
\end{equation}
the diurnal thermal skin depth:
\begin{equation}
    L \equiv \sqrt{2K \over \omega \rho C} \,,
\end{equation}
where $\omega$ is the angular frequency of the asteroid, and the thermal parameter:
\begin{equation}
    \Theta \equiv { \sqrt{K \omega \rho C } \over 4\sqrt{2}\pi^{-{3\over 4}}\epsilon\sigma u_\star^3} \,.
\end{equation}
Numerical factors in these definitions arise from the derivation of the analytical solution (see Appendix~\ref{sec:analytic}); however, some authors do not use them \citep{Lagerros_1996, Golubov_2012}.

If we neglect self-heating terms, the heat diffusion equation (\ref{HDE}) and 
its boundary condition (\ref{BC}) can be rewritten using dimensionless variables 
$\vec{\xi} \equiv \vec{r} / L $, 
$\varphi \equiv \omega t$, 
$\upsilon \equiv u /  u_\star$ as:
\begin{align}
    {1 \over 2}\Delta_{\xi}\upsilon - {\partial \upsilon \over \partial \varphi} &= 0 \,,\\
    4\pi^{-{3\over 4}}\Theta \vec{n}\cdot \nabla_{\xi} \upsilon + \upsilon^4 &= \vec{s} \cdot \vec{n}  \,,
\end{align}
where $\nabla_\xi$, $\Delta_\xi$ is the gradient and 
the Laplacian with respect to the variable $\xi$.  
The only independent parameter in these equations is the thermal parameter $\Theta$. 
However, the boundary condition must hold for all $L \vec{\xi} \in \partial \Omega$. 
If $\ell$ is the characteristic size of the boulder, 
then the problem of finding a~dimensionless temperature $\upsilon$ 
has actually two independent parameters --- %
the thermal parameter $\Theta$ and the dimensionless size $\ell / L$. 
In the following, we select the characteristic size as 
    the  square root of the base area of the boulder, 
$\ell \equiv \sqrt{S}$.


\subsection{The dimensionless pressure}
\label{sec:pressure}

\begin{figure}
    \centering
    \includegraphics[width=0.3\textwidth]{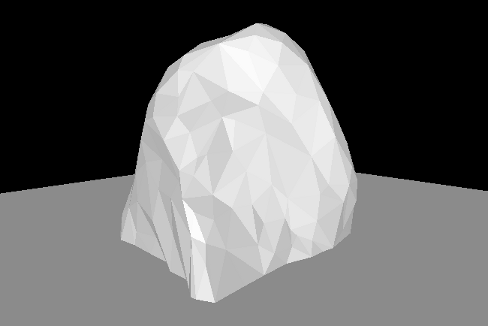}%
    \caption{An example of the boulder shape which was used to compute the surface temperature distribution. }
    \label{img:boulder_img}
\end{figure}

In our model, the shape of the boulder can be arbitrary. 
As a~special case, we can choose a~high wall and 
compare our results to the model of \cite{Golubov_2012}. 
We started with such an idealized ``boulder'' to determine 
the influence of the self-heating effect and 
the absorption of radiation on the recoil force  
(see Appendix~\ref{sec:comparison}).
Nevertheless, we then selected a~boulder of \emph{realistic} irregular shape 
for our computations, as shown in Figure \ref{img:boulder_img}. 
The shape was obtained by a 3D scanning of a randomly selected boulder.
In this case, we do not take into account the self-heating effect nor the influence of absorption though; this decision is justified in Appendix~\ref{sec:comparison} as well.

\begin{figure}
    \centering
    \includegraphics[width=0.3\textwidth]{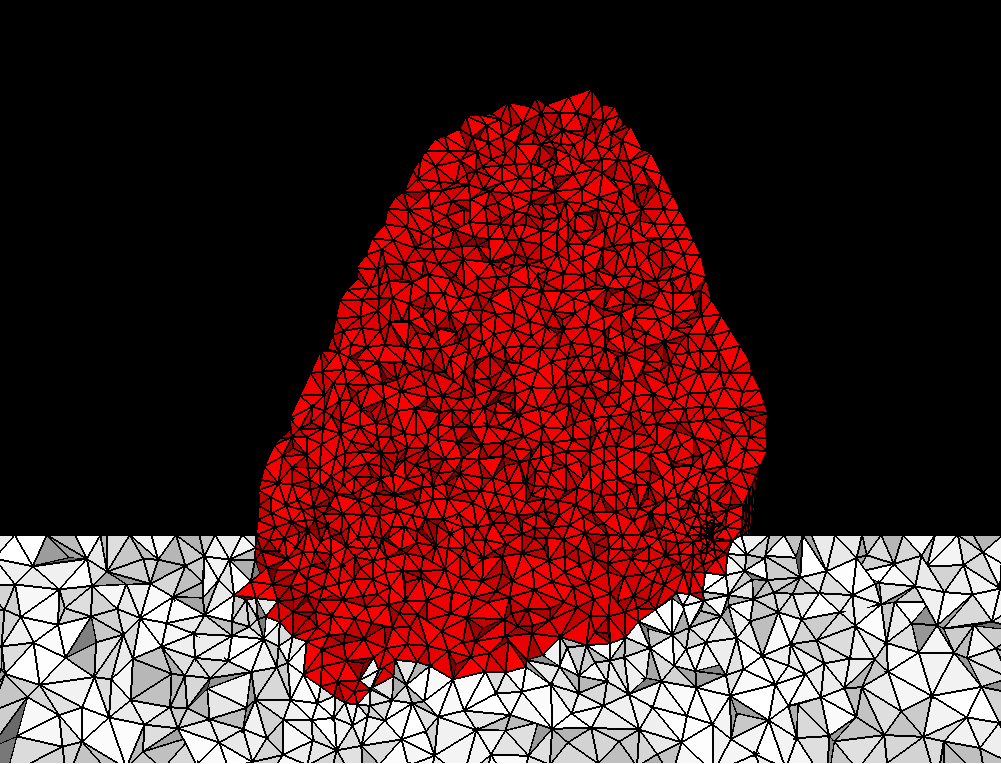}
    \caption{A sectional view of the 3D mesh, created by the Tetgen code.
        Thermal parameters of the boulder are varied in our model. Nonetheless, we assume the boulder has generally different thermal conductivity than the surrounding layer of regolith. 
    Thermal properties of the regolith as given by \citet{Farinella_1998} are as follows: the density $\rho = 1500 \,\rm kg\,m^{-3}$, the thermal conductivity $K =  0.0015 \,\rm W\,m^{-1}\,K^{-1}$ and the specific heat capacity $C = 680\,\rm J\,kg^{-1}\,K^{-1}$. We chose the hemispherical albedo $A=0.1$, the infrared emissivity $\epsilon = 0.9$, and the rotational period equal to $P = 12.1 \,$hours. }
    \label{img:domain_parts}
    \end{figure}

    We considered different values of the thermal conductivity for the studied boulder and for the surrounding layer of regolith, 
as demonstrated in Figure \ref{img:domain_parts}. 
Thermal properties of the regolith are taken from \cite{Farinella_1998}, 
properties of the boulder are determined by values of thermal parameter~$\Theta$ and the skin depth~$L$. 

We should stress the importance of the \emph{non-linearity} of the problem. 
We derived a~linearized analytical solution of the heat diffusion equation in a~half-space domain (see Appendix~\ref{sec:analytic}), 
where we deal with the non-linear term $u^4$ by substituting $u_0^4 + 4u_0^3 \delta u$, 
where $u_0$ is a~constant, 
$\delta u$ is the change of temperature. 
The same term appears in the expression for the recoil force (\ref{f_rad}) from a~thermal emission. 
In case of a~symmetric boulder, 
the complete linearization would lead to identically zero mean torque. 
Therefore, we solve a~non-linear problem by an iterative method, as described in Section~\ref{sec:HDE}.

The solution of the HDE is a~time-dependent temperature distribution in the boulder, 
particularly on its surface. 
It follows that we can determine the recoil force the boulder exerts; 
the force is given by the formula (\ref{f_rad}).
However, it is convenient to introduce the dimensionless pressure:
\begin{equation}
\Pi \equiv {2 \over 3} {1 \over S} \int_{\Gamma_1} {u^4 \over u^4_\star} n_y\, \boldd\Gamma \,,
\label{Pi}
\end{equation}
where $n_y$ is the $y$-th component of the local normal, 
$S$ is the base area of the boulder. 
The dimensionless pressure allows us to compare the magnitude of the tangential force for different sizes of the boulder. 

The projection of the total torque to the rotational axis is then given by:
\begin{equation}
    \vec \tau \cdot \vec{e} = { (1-A)\Phi \over c} \Pi S\, r_\perp  \,,
\end{equation}
where $r_\perp$ is the distance of the boulder from the rotational axis.

If we consider a~wall aligned with a~local meridian, which face of area $S$ has a~constant temperature $u$, our definition (\ref{Pi}) is then reduced to:
\begin{equation}
    \Pi = {2 \over 3}{u^4 \over u^4_\star}\,,
\end{equation}
which is equivalent to the definition of a~dimensionless pressure by \cite{Golubov_2012}. Our definition is therefore analogous, with the exception of the area $S$ being the projection to the plane $xy$ rather than the plane $xz$.

The dimensionless pressure varies during a~rotation. 
We can obtain a~measure of the long-term effect by averaging over one rotational period; 
to this point we introduce the mean dimensionless pressure:
\begin{equation}
    \langle \Pi \rangle = {1 \over P} \int_0^P\!\! \Pi\,\boldd t \,.
\end{equation}

\begin{figure}
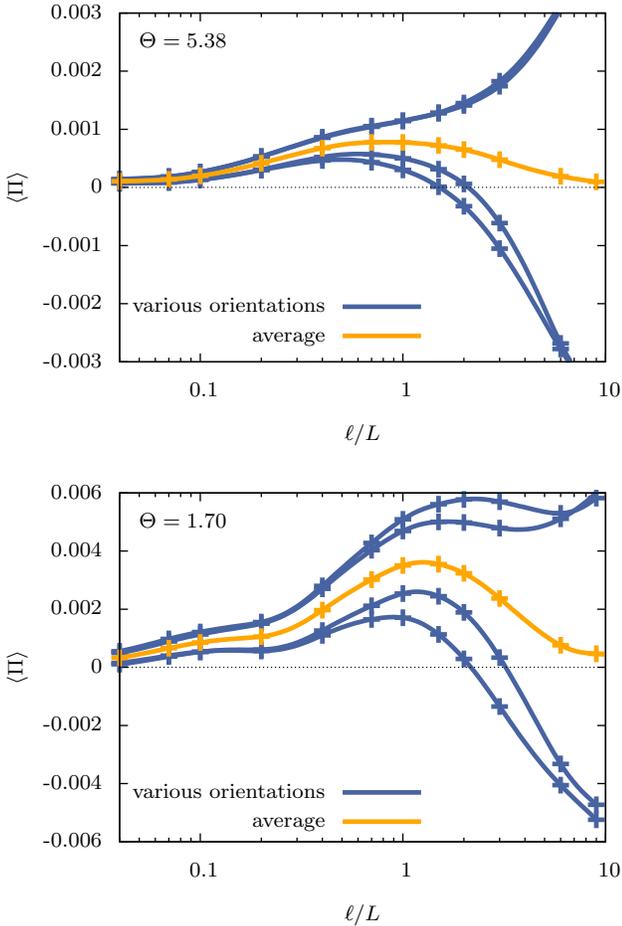

    \centering
    \input boulder_table_rev \par
    \input boulder_table_rev_2
    \caption{Computed values of the mean dimensionless pressure~$\langle \Pi \rangle$ 
        as a~function of the dimensionless boulder size $\ell / L$ 
        for two different values of the thermal parameter 
            $\Theta = 5.38$ and $1.70$ 
        (notice that vertical axes have different scalings). 
        The dark curves correspond to the boulder rotated by 
        $0^\circ$, $90^\circ$, $180^\circ$ and $270^\circ$ 
    around the vertical axis. 
        We notice that all curves approach a~zero for $\ell / L \rightarrow 0$ 
    and they exhibit a~maximum for $\ell \sim L$.
        For $\Theta = 1.70$ there seems to be an inflection at about 
        $\ell \sim 0.1\,L$, which is absent for $\Theta = 5.38$.
        The bright curve is the arithmetic mean of the dark curves. 
        We assumed a simpler ``shadowing'' model in this case 
        (but cf. Figure~\ref{img:wall}).}
    \label{img:boulder_mdp}
\end{figure}


\subsection{The total pressure exerted by a set of variously oriented boulders }
\label{sec:orientations}

The mean dimensionless pressure $\langle \Pi \rangle$ as a~function of the dimensionless size $\ell / L$ varies significantly for different shapes of the boulder, or even for different orientations of the same boulder. 
It is evident that the limit of very high conductivity (that is $\ell / L \rightarrow 0$) leads to a~zero dimensionless pressure~$\Pi$ for \emph{all} shapes of a~boulder. 
In such a~case, the boulder is isothermal and therefore emits the same flux to the western and eastern directions, resulting in a~null torque. 

The limit of the mean dimensionless pressure for zero thermal conductivity ($\ell / L \rightarrow \infty$) differs from boulder to boulder. The conductive term in the energy balance equation (\ref{BC}) is negligible and the temperature at a~given point of surface is determined by the immediate balance between incoming and outgoing radiant flux.
Since we solve the HDE in a~single boulder only, we need to obtain a~torque (as a~function of a~boulder size) that would represent a~set of all boulders on the surface. 
It is reasonable to assume that the boulders are randomly oriented on the surface. Although some orientations of boulders seems to be preferred on certain parts of the surface of Itokawa \citep{Miyamoto_2007}, we anticipate that no orientation prevails on a~global scale.
The total torque induced by boulders will therefore vanish in the limit of zero conductivity.
For that reason, we demand the mean torque $\langle \Pi \rangle$ to approach zero as well.
However, in the general case of an asymmetric boulder, the mean dimensionless pressure will approach a~non-zero value. 

We have several options how to resolve this issue.
For instance, we can restrict ourselves to symmetric boulders only. If the boulder is symmetric with respect to the plane of the local meridian, the mean pressure will vanish in the limit case. Nonetheless, we want to keep the universality of our model and use irregular asymmetric boulders. 
In this case, we can compute the mean pressure $\langle \Pi \rangle$ for several orientations of the boulder and then take the average of these values. Another possibility is to calculate the mean pressure for a~single orientation and subtract the pressure in the limit of zero conductivity. We employed the former option. 

Figure  \ref{img:boulder_mdp} shows the mean dimensionless pressure for several orientations of the studied boulder and the averaged values. 
We assumed the boulder lies on the equator of an asteroid.
We see that the averaged values approach zero in the limit of zero conductivity, as expected. 


\subsection{A dependence of the mean pressure on asteroidal latitude } \label{sec:latitude}

For an asteroid with zero obliquity, the body--Sun vector $\vec{s}$ has Cartesian coordinates:
\begin{equation}
    \vec{s} = (\sin \vartheta \cos \rm HA, -\sin \rm HA, -\cos \vartheta \cos \rm HA) \,,
\end{equation}
where $\rm HA$ is the hour angle and $\vartheta$ is the asteroidal latitude (defined as $\sin \vartheta  = \vec e \cdot \vec n$).
 The dependence of the dimensionless pressure $\Pi$ on the hour angle $\rm HA$ vanishes after averaging over a~rotational period, the dependence on $\vartheta$ remains.
 
Assuming we can separate variables $\ell$, $\vartheta$, we can decompose the mean pressure $\langle \Pi \rangle$ as:
\begin{equation}
    \langle\Pi \rangle(\ell, \vartheta) = \mathcal{P}(\ell)\mathcal{V}(\vartheta) \,,
\end{equation}
where $\mathcal{P}(\ell) = \langle \Pi \rangle (\ell, 0)$. The function $\mathcal{V}(\vartheta)$ constitutes a~latitudinal dependence and is normalized such that $\mathcal{V}(0^\circ)=1$. It obviously depends on the shape of a~boulder. For this test, we chose an approximately hemispherical boulder, because it is axially symmetric and therefore one latitude is not preferred over other values. 

\begin{figure}
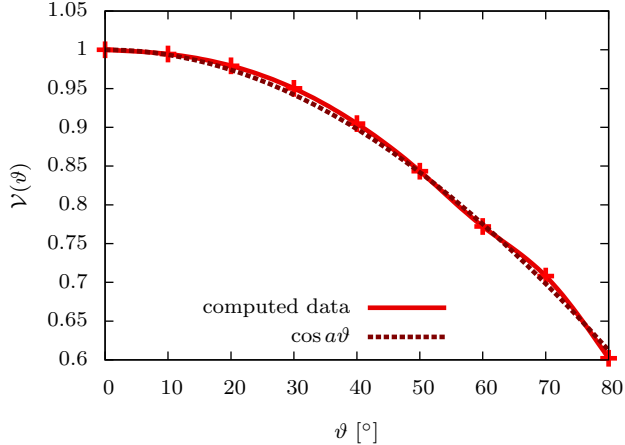

    \centering
    \input hemi_theta
    \caption{The dependence of the mean dimensionless pressure~$\langle \Pi \rangle$ on the asteroidal latitude~$\vartheta$ of the boulder.}
    \label{img:hemi_theta}
\end{figure}

We show the computed values of the function $\mathcal{V}(\vartheta)$ in Figure~\ref{img:hemi_theta}. 
It can be approximated by a~function $\cos a\vartheta$, 
where $a = 0.653\pm0.004$ is a~parameter determined by a~least square fit. 
The mean pressure $\langle \Pi \rangle$ is maximal for the boulders located on the equator
and does not drop below 50\% of the maximum for $\vartheta = 80^\circ$ latitude.
The known dependency of the mean pressure on the latitude allows us to compute the torque induced by a~boulder on any given point of the surface, 
which we shall use in the next section.

\subsection{An influence of the elliptical trajectory}
\label{sec:eccentricity}
\begin{figure}
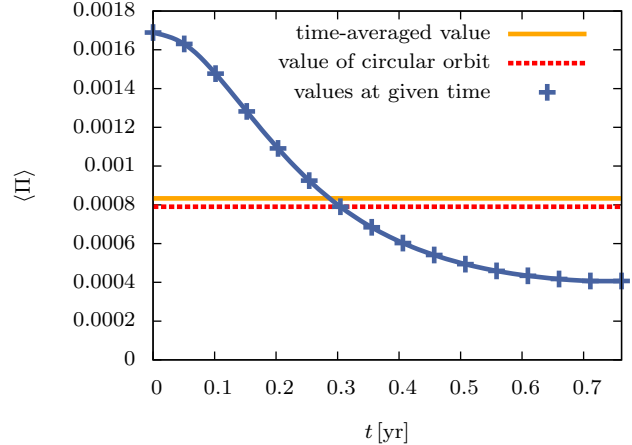

    \centering
    \input orbit
    \caption{\chng{The values of the mean dimensionless pressure $\langle\Pi\rangle$ 
        from the perihelion ($t=0$) to aphelion ($t=P/2$). The values are computed 
        for the high-conductivity case ($\Theta=5.38$ at the distance of the semimajor axis) 
        and for $\ell = L$, averaged over four different orientations
        of the boulder. One can readily see 
    that the difference between the time-averaged value and the value of corresponding 
    circular orbit is negligible, given other uncertainties of our model.} }
    \label{img:orbit}
\end{figure}

\chng{%
So far we assumed the asteroid orbits on a circular trajectory.
This assumption allowed us to ignore any seasonal effects 
and average the torque over a rotational period only. However, the 
eccentricity of Itokawa is $e=0.28$; it is not clear whether the ellipticity 
of the trajectory affects our model significantly. We thus compute the mean 
dimensionless pressure $\langle \Pi\rangle$ in several ``discrete'' points 
on the orbit separated by a time step equal to $1/30$ of the orbital period. 
We assume a high-conductivity case here. Nevertheless,
the thermal parameter $\Theta$ will vary over time;
it depends on the distance from the Sun as $\Theta \sim r^{3\over2}$,
so it will reach $\Theta=3.29$ in the perihelion and $\Theta = 7.79$ in the aphelion.
We restrict ourselves to a single value of the dimensionless size $\ell/L = 1$
and we average the result over four different orientations of the boulder, as before.
Then, we obtain the time-averaged value by simply taking an arithmetic mean of the computed values. 

This approximate average over elliptical orbit is to be compared with the mean dimensionless pressure 
corresponding to the circular orbit, which we obtained already in Section \ref{sec:orientations}.
The result can be seen in Figure~\ref{img:orbit}.
We see that even for the considerable eccentricity of Itokawa, the difference seems 
negligible. The value differs from our previous result by less than $5\,\%$.
Therefore, the approximation of the circular trajectory is well justified.
}

\section{The angular acceleration of asteroid (25143) Itokawa}\label{sec:itokawa}

In the following, we focus our attention on the asteroid (25143) Itokawa. First, we estimate the total number of boulders on the surface and then compute how the thermal emission from boulders alters the angular acceleration predicted by global-shape models of the YORP effect. 

Existing models of the YORP effect usually assume a~normal direction of the recoil force. 
However, for non-convex asteroids the force can be influenced by the absorption of radiation emitted by the surface \citep{Statler_2009}.
In the previous section, we demonstrated that a~surface feature can alter the recoil force as well. 
We pointed out that the lateral heat diffusion through boulders leads to an emergence of the \emph{tangential} component of the recoil force. 
The presence of surface features also changes the normal component. 
While a~complete solution would require solving the heat diffusion equation in the \emph{whole} asteroid, including boulders, we neglect the change in the normal component and we solve for the tangential component separately.

The torque generated by a~single boulder was discussed in previous chapter \ref{sec:Pi}. We now place a~large number of such boulders on the shape model of Itokawa and calculate the torque they induce. The total YORP torque and corresponding change of the angular velocity of the asteroid is then obtained by adding our result to the result of the global-shape model of the YORP effect.


\subsection{The torque induced by boulders}

We demonstrated the emergence of the \emph{tangential} component of the force, 
which is of our interest.
Therefore, we consider a~direction of the force perpendicular to the local normal $\vec n$, 
regardless of the location on the surface of the asteroid. 
Although the direction depends on the shape, we discussed that the overall 
effect corresponds to the force perpendicular to the rotational axis $\vec e$. 
In follows that the direction of force is $\vec e \times \vec n$. 

The magnitude of $\vec e \times \vec n$ is proportional to $\cos \vartheta$. 
However, we studied the dependence of the torque on asteroidal latitude in Section~\ref{sec:latitude}. 
We showed that the mean torque as a~function of latitude $\vartheta$ is approximately $\cos 0.653\vartheta$.
The recoil force induced by single boulder is therefore proportional to the term:
\begin{equation}
\vec f \sim {\cos 0.653 \vartheta \over \cos \vartheta}  \,\vec e \times \vec n \,.
\end{equation}

Our goal is to compute the total torque $\vec \tau$ induced by the boulders; 
to be more precise, its projection $\vec \tau \cdot \vec e$ to the rotational axis.
Let $N(\ell) \boldd \ell$ be the number of boulders in the size interval $(\ell, \ell+\boldd \ell)$. 
We can write the size distribution~$N(\ell)$ as:
\begin{equation}
N (\ell) \boldd \ell = N_0 \varrho(\ell) \boldd \ell\,,
\label{prob_dens}
\end{equation}
where 
$N_0$ is the total number of boulders on the surface and 
$\varrho(\ell)\boldd \ell$ is the probability that a~randomly selected boulder has a~size in the interval $(\ell, \ell+\boldd \ell)$.
We compute the total torque induced by boulder as follows. 
We select a~facet of Itokawa shape model randomly. 
Assuming boulders cover Itokawa uniformly, the probability of selecting each facet is proportional to its area. 
Now, we generate a~random size~$\ell$ with the probability distribution~$\varrho$, using the inverse sampling method. 
The torque induced by the boulder on the selected facet is added to the total torque. 
The final number of generated boulders is given by the value $N_0$ from Eq.~(\ref{prob_dens}). 
Because the number of boulders is very high, the total torque basically does not depend 
on the realization of boulder distribution on the surface.  

It is clear that the boulder size distribution $N(\ell) \boldd \ell$ constitutes 
an important parameter for a~calculation of the total torque. 
We thus discuss the size distribution of boulders on Itokawa in the following section.


\subsection{The observed size distribution of small boulders}
\label{sec:size_dist}

In order to obtain the torque caused by boulders, 
it is necessary to find out the total number of boulders and their size distribution. 
The differential size distribution of boulders larger than $5\, \rm m$ 
on the whole surface of Itokawa can be approximated by a power law \citep{Saito_2006}:
\begin{equation}
    N(\ell) \boldd \ell \approx 1.3 \times 10^5 \, [\ell]_{\rm m}^{-3.8} \boldd \ell \,.
    \label{size_dist_large}
\end{equation}
Surface images taken by the Hayabusa spacecraft revealed that this power law
can be extrapolated down to sizes of $0.1 \, \rm m$  on certain parts of the surface \citep{Miyamoto_2007}, 
although the slope of a~log-log graph falls off significantly for smaller sizes.
However, other parts of the surface clearly show a~different topography. 
Furthermore, an extrapolation of the above-mentioned size distribution down to $1 \, \rm mm$ is clearly unacceptable;
boulders of sizes between $1\, \rm mm$ and $0.1 \, \rm m$ alone would cover about $4 \times 10^{7}\,\rm m^2$ of the surface area, 
but the surface of Itokawa is only $3.93 \times 10^5 \,\rm m^2$ \citep{Demura_2006}. 

\begin{figure}
    \centering
    \includegraphics[width=0.4\textwidth]{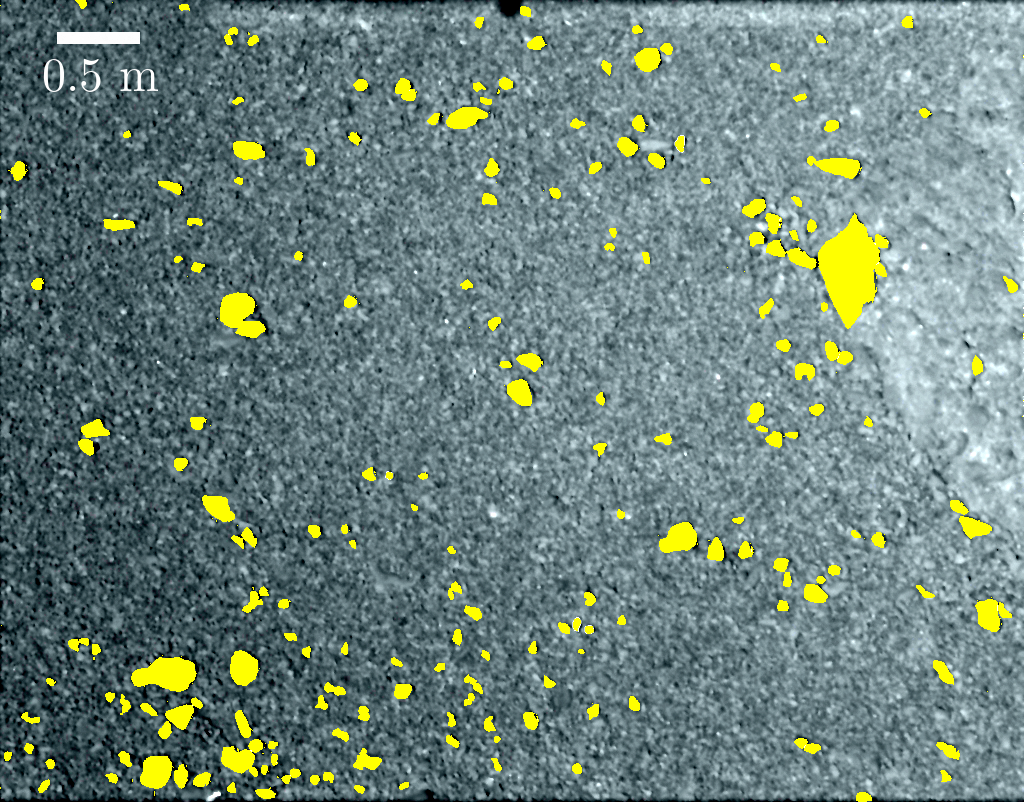}
    \caption{The image ST\_2563607030\_v \citep{Saito_etal_2010} with highlighted boulders from which we derived their size distribution, used for the computation of the total torque.}
    \label{img:boulders_photo}
\end{figure}

\begin{figure}
    \centering
    \input{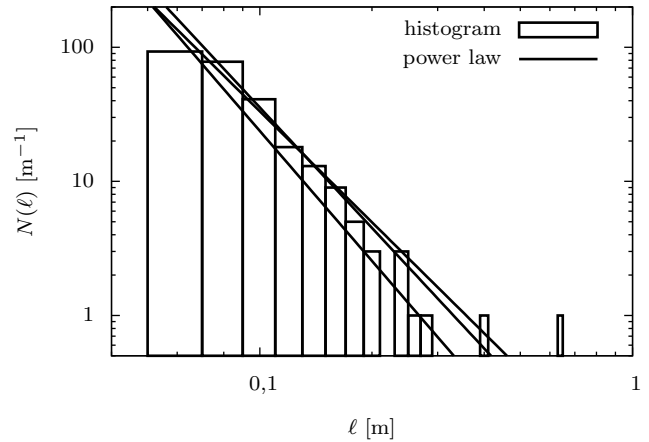}
    \caption{The histogram of small boulder sizes on the surface of Itokawa, constructed from images ST\_2563537820\_v and ST\_2563607030\_v. 
    The three power-law fits correspond to the average fit and 
        left/right 1-$\sigma$ uncertainty, determined by a Monte-Carlo method.
        Two rightmost bins lie off the linear part of the histogram and 
        were not included in the least-square method dataset.
        The values of the multiplicative constant and the exponent of the power law
        are shown in Eq.~(\ref{size_dist_small}).}%
        \label{img:hist_dist}
\end{figure}

Therefore, we sought for a~different size distribution of small pebbles. 
We estimated their size distribution from close-up images taken by the Hayabusa during its descend, 
namely the images ST\_2563537820\_v and ST\_2563607030\_v from \cite{Saito_etal_2010} dataset.
The resolution of these images is $7\,\rm mm/pixel$ and $6\,\rm mm/pixel$, respectively \citep{Miyamoto_2007}, 
which allows us to find distinct boulders only few centimeters in size. Identified boulders are shown in Figure~\ref{img:boulders_photo}.
We constructed a~histogram of sizes (see Figure \ref{img:hist_dist}) from which we inferred the size distribution:
\begin{equation}
    N(\ell)\,\boldd\ell = (14\pm 9)\times 10^3\, [\ell]_{\rm m}^{-\left(3.0\pm 0.2\right)}\,\boldd \ell\,.
    \label{size_dist_small}
\end{equation}%
We assume this power law can be extrapolated to millimeter-sized pebbles. 
Considering the uncertainties, the area covered by boulders of sizes between $1\,\rm mm$ and $1\,\rm m$ would be between 19\,\% and 32\,\% of the total surface area. 
This finding seems to be viable. 

\subsection{A comparison of the YORP torque by boulders and by the global shape} 

The global-shape YORP effect model of the asteroid Itokawa predicts a~significant rotational deceleration, 
which is inconsistent with the observed acceleration. 
As mentioned above, the lateral heat diffusion through boulders induces an additional torque, 
which affects the angular velocity. 

Let the magnitude of the total torque generated by boulders be $\tau$. 
The asteroid will undergo the rotational acceleration:
\begin{equation}
    {\boldd \omega \over \boldd t} = \left({\boldd \omega \over \boldd t}\right)_{\rm global} \!\!+ {\tau \over I} \,,
\end{equation}
where $(\boldd\omega/\boldd t)_{\rm global}$ is the prediction of the global-shape YORP model, $I \doteq 7.77 \times 10^{14} \rm \,kg\,m^2$ is the moment of inertia of Itokawa \citep{Scheeres_2007}. 

\begin{figure}
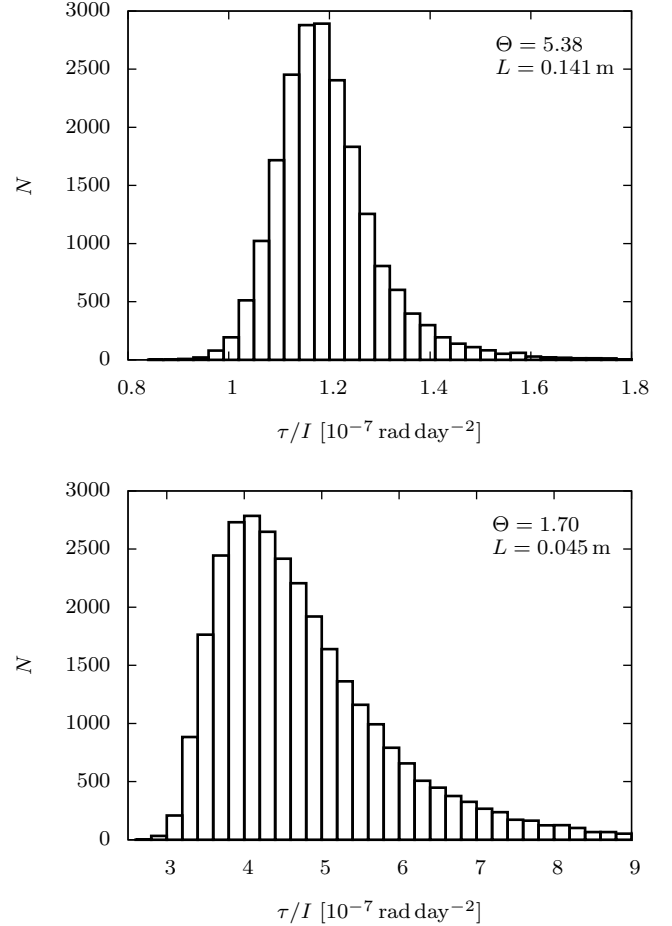

    \centering
    \input dwdt_hist\par
    \input dwdt_hist_2
    \caption{The contribution of boulders to the total 
    angular acceleration of Itokawa,
    computed for two different values 
    of the thermal parameter~$\Theta=5.38$ and $1.70$ and 
    the corresponding thermal skin depth~$L$.
    Each histogram was constructed by a Monte-Carlo method;
    individual runs correspond to different power laws 
    of the boulder size distributions. 
    Note that horizontal axes differ in each figure.
    Eqs.~(\ref{T/I}) and (\ref{T/I2}) show the average value 
    and 1-$\sigma$ uncertainties.  }
    \label{img:dwdt_hist}
\end{figure}

The global-shape model of the YORP effect predicts a~rotational deceleration $-(2{\rm\ to\ }6) \times 10^{-7}\,\rm rad\,day^{-2}$, 
depending on the resolution of the shape model \citep{Breiter_2009}. 
In order to determine the torque induced by boulders, 
it is necessary to select values of the parameters --- namely the thermal parameter~$\Theta$ and the thermal skin depth~$L$.
We adopted following material properties: $K = 2.65 \,\rm W\,m^{-1}\,K^{-1}$, $C = 680\,\rm J\,kg^{-1}\,K^{-1}$, $\rho = 2700 \,\rm kg\,m^{-3}$. 
Together with orbital parameters of Itokawa, this yields the thermal parameter 
$\Theta = 5.38$ and the thermal skin depth $L=0.141\,\rm m$. 
Utilizing the size distribution of boulders derived in Section~\ref{sec:size_dist}, 
we obtain the result:
\begin{equation}
    \left.{\tau  \over I}\right|_{\Theta=5.38} = (1.20\pm 0.11) \times 10^{-7} \,\rm rad\,day^{-2} \,.
    \label{T/I}
\end{equation}
As the thermal conductivity seems to be the most uncertain parameter, 
     we also computed the torque for a lower value, 
$K=0.26\,\rm W\,m^{-1}\,K^{-1}$ 
(keeping other parameters unchanged). 
In such a case, the thermal parameter is $\Theta = 1.70$ and 
the thermal skin depth $L=0.0446\,\rm m$.
The corresponding torque is then:
\begin{equation}
    \left.{\tau\over I}\right|_{\Theta=1.70} = (4.8 \pm 1.2) \times 10^{-7}\,\rm rad\,day^{-2}\,.
    \label{T/I2}
\end{equation}
The probability distribution for both cases 
    is shown in Figure \ref{img:dwdt_hist}.
    \par
For the sake of comparison, we can refer to the result of global-shape models 
of \cite{Lowry_2014}, $(-1.80\pm 1.96)\times 10^{-7} \,\rm rad\,day^{-2}$,
or \cite{Breiter_2009}, $-(2.5 \rm\ to \ 5.5) \times 10^{-7} \,\rm rad\,day^{-2}$.
We notice that this torque is of the same order as our result,
but has an opposite sign. 
The torque induced by boulders and the torque from the global asymmetry could effectively cancel out, 
resulting in the change of angular velocity much smaller than predicted by global-shape models.
As the observed angular acceleration of Itokawa is $(0.35 \pm 0.04) \times 10^{-7} \, \rm rad\,day^{-2}$ \citep{Lowry_2014}, 
our model presents an alternative explanation of the observed acceleration 
without any need for a~non-uniform density distribution.

\section{Conclusions}\label{sec:conclusions}

In this paper, we presented a~detailed numerical model of the local YORP effect induced by a~boulder or a~set of boulders.
The three-dimensional heat diffusion equation in the boulder was solved using the finite element method. 
Unlike the finite difference method, the finite elements have basically no restriction on the shape of a~domain, 
allowing us to solve the heat diffusion equation in the boulder of a~realistic shape. 
Furthermore, we assumed the studied boulder has a~different thermal conductivity than the surrounding regolith layer. 

Our boulder had a~general asymmetric shape, so it exhibited a~non-zero torque even in the limit of zero thermal conductivity. 
However, this torque depends on the orientation of the boulder. 
In order to obtain an~average torque representing a~set of randomly-oriented boulders, 
we computed torques for several orientations and then the average of these values. 
We verified that the averaged torque approaches zero in the zero conductivity limit.

The non-zero torque arises from the asymmetry of the emission (averaged over the rotational period). 
There are two reasons for the emission asymmetry. 
The first is the asymmetry of the boulder shape. 
Indeed, \cite{Rubincam_2000} demonstrated an emergence of the YORP effect on a~toy model of a~spherical asteroid with two wedges attached to its equator. 
The torque created by the emission from the vertical side is greater in magnitude than the torque created by the emission from the inclined side, 
thus resulting in a~non-zero total torque.

The second reason for the emission asymmetry is the lateral heat diffusion through the boulder
(as in \citealt{Golubov_2012}).
We can imagine a~boulder on the equator. 
In the morning, the eastern side of the boulder is heated up and the boulder exerts a~recoil force of western direction.
Provided the width of the boulder is comparable to the thermal skin depth, 
the heat diffusion contributes to heating of the western side in the afternoon. 
The emission from the western side is therefore more intense. 
The recoil force has an eastern direction and exceeds the force from the eastern side in magnitude, 
thus creating a~non-zero mean force of eastern direction. 
The corresponding torque causes an angular acceleration of the asteroid.

The global contribution of the shape asymmetry of boulders to the YORP effect is likely to be null, 
because of the very large number of boulders on the surface. 
In contrast, the lateral heat diffusion leads to the torque with a~direction of the rotational axis, 
thus accumulating over individual boulders. 
Even though the torque generated by a~single boulder is tiny, 
the overall effect can be comparable to the global-shape YORP effect, 
if there is a~sufficient amount of boulders, of course.

We showed the maximum pressure is exerted by boulders of sizes comparable to the diurnal thermal skin depth~$L$. 
Figure \ref{img:boulder_mdp} shows the dependence of the pressure on the size of the boulder. 
However, this graph does not reflect the actual contribution of boulders of different sizes to the total torque. The decisive factor is the exponent $\gamma$ of the power-law size distribution. If $\gamma>2$, smaller boulders will exert a higher torque (compared to the Fig. \ref{img:boulder_mdp}). In our case, the overall torque is generated mostly by boulders of sizes between $0.1\,L$ and $L$.

The general approach allowed us to compare our three-dimensional model with the one-dimensional model of \cite{Golubov_2012}. 
In case of a~\emph{symmetric} boulder, 
we confirmed that the torque vanishes in the limits of high conductivity and zero conductivity, 
We showed that the maximum torque
appears for $\ell \sim L$.  

Unlike \cite{Golubov_2012}, we found positive values of the torque for all parameter ranges in the case of a~symmetric boulder. 
An asymmetric boulder could produce a~negative torque, 
but after averaging over orientations the resulting torque is again strictly positive. 
Even \cite{Golubov_2012} realized the torque is mostly positive, 
and proposed a~possibility of an equilibrium between the global-shape torque and the torque induced by boulders, 
resulting in a~null total torque; 
they suggested this could be the case of the asteroid (25143) Itokawa. 
However, \cite{Lowry_2014} detected a positive change in angular velocity of Itokawa, 
which means this asteroid is not in such equilibrium state.

Of course, our model contains a~number of free parameters that can change the magnitude of the torque significantly. 
The crucial factor is the total number of boulders and their size distribution. 
We realized that the size distribution of large boulders on Itokawa of \cite{Saito_2006} cannot be extrapolated to centimeter sizes. 
Having no better alternative, we estimated the size distribution from close-up images of the surface of Itokawa 
and we extrapolated it for sizes of indiscernible pebbles. 
We also assumed that other parts of the surface have the same size distribution of boulders. 
If we neglect an interaction between boulders (mutual shadowing or thermal irradiation), which seems to be reasonable given the separations of boulders seen on Figure \ref{img:boulders_photo}, 
the torque is directly proportional to the number of boulders. 

The choice of the lower limit of the size distribution might be also a~disputable parameter. 
Although the magnitude of the torque induced by a~boulder approaches zero as the size of the boulder approaches zero, 
even sub-millimeter pebbles could have a~non-negligible influence on the total torque. 
However, it is doubtful whether so small particles can be considered as boulders, 
or whether they form a~uniform layer of matter. 
We selected the lower limit $1\,\rm mm$. 

The shape of the studied boulder is another key factor of our model. 
We selected a~boulder of a~realistic irregular shape, but it was selected ad hoc. 
There are two particularly important properties of boulders:
 their height and flatness. 
The higher the boulder is, the greater torque it likely induces. 
If the sides of the boulder are perpendicular to the surface, 
the lever arm of the torque is maximal; 
the lower the slope of sides, the lower lever arm.

We finally applied our model to the case of the asteroid (25143) Itokawa.
We showed that boulders could induce a~torque that would cause 
the angular acceleration of the order  $10^{-7} \,\rm rad\,day^{-2}$. 
We realized there was a~significant uncertainty; 
nevertheless, we clearly demonstrated that the emission from boulders \emph{is capable} of producing the torque 
comparable in magnitude to the global-shape YORP effect.
Our result is consistent with the observed acceleration \citep{Lowry_2014} 
and presents an alternative and viable explanation of the discrepancy between the observed acceleration and the acceleration predicted by global-shape models.

\section{Future work}
We postpone the following topics for future work. 
We assumed a~rotational axis perpendicular to the orbital plane, 
therefore we need not consider the orbital movement and the torque is averaged over the rotational period only. 
Luckily, the obliquity of Itokawa is approximately $178^\circ$ \citep{Demura_2006}, 
which is very close to the perpendicular state. 
Considering a~general direction of the rotational axis, 
the insolation will change during the revolution about the Sun and the seasonal changes of temperature will occur. 
We expect a seasonal variant of the studied effect to appear on boulders whose size is comparable to the \emph{seasonal} thermal skin depth. 

We already discussed that the diurnal torque has a~direction of the rotational axis, 
as it is caused by the asymmetry of emission between the western and the eastern part of the boulder due to the lateral heat diffusion. 
Following the same principle, the seasonal torque could be caused by the asymmetry between the northern and the southern part of the boulder. 
The direction of this torque would therefore be perpendicular to the rotational axis. 
We thus anticipate the seasonal component of the effect will not affect the angular velocity of the asteroid, 
it will only cause an evolution of the obliquity. 

In our model, we assumed a shadow is casted by the boulder; however, 
we did not take into account shadows casted by global-shape inconvexities 
of Itokawa. The same goes for the self-heating effect, which is also
considered only locally. To resolve this issue, it would be necessary to determine
the insolation function for each facet of the shape model separately
and then solve many three-dimensional heat diffusion problems.


\section{Acknowledgements}

The work of MB has been supported by the Grant Agency of the Czech Republic
(grant no.\ 13-01308S), and the work of JD by the GACR,
grant no.\ P209/10/0537.
We thank the first referee S.\ C.\ Lowry for many valuable suggestions
and the second (anonymous) referee for pointing out important corrections.


\bibliographystyle{elsarticle-harv}
{\scriptsize
\bibliography{references}
}


\appendix

\section[]{A linearized analytical solution of the one-dimensional heat diffusion equation}
\label{sec:analytic}

The general three-dimensional heat diffusion equation with a~non-linear boundary condition in an irregular domain has no analytical solution. 
To find the temperature distribution, we must employ a~numerical approach such as the finite element method. 
Nevertheless, it is useful to derive an analytical solution for a~simplified case of a~half-space domain, 
which allows to reduce the problem to one spatial dimension only, as in \cite{Capek_2007}.
The solution can be used as a~test for our numerical model and also as a~Dirichlet boundary condition (see Figure~\ref{img:domain}).

Suppose the Sun illuminates an (infinite) plane $z=0$ and a half-space $z>0$ represents the domain $\Omega$. 
We seek for the temperature $u$ as a~function of the depth~$z$ and time~$t$, solving the heat diffusion equation:
\begin{equation}
    \alpha\,{\partial^2_z u} -{\partial_t u } = 0 \,,
    \label{HDE_1D}
\end{equation}
with boundary conditions:
\begin{align}
    \label{HDE_boundary_1} -K{\partial_z u}(0,t) + \epsilon \sigma u^4(0,t) &= \mathcal{F}(t) \,, \\
    \label{HDE_boundary_2} {\partial_z u }(\infty, t) &= 0 \,,
\end{align}
where $\alpha \equiv {K \over \rho C}$~is the thermal diffusivity, 
$K$~the thermal conductivity, 
$\rho$~the density, 
$C$~the specific heat capacity, 
$\epsilon$~the infrared emissivity, 
$\sigma$~the Stefan-Boltzmann constant 
and $\mathcal{F}(t)$~the incoming radiant flux. 
The first boundary condition is the energy balance equation 
and the second one is necessary for a uniqueness of the solution --- 
it eliminates solutions where the temperature rises ad infinitum as $z \rightarrow \infty$.

The radiant flux is a~periodic function, therefore we can represent it as a~Fourier series,  $\mathcal{F}(t) = \sum_{n=-\infty}^\infty \mathcal{F}_n e^{\i n\omega t}$. 
We look for a~\emph{stationary} solution, which we can represent by a sum $u(z,t) = \sum_{n=-\infty}^\infty u_n(z) e^{\i n \omega t}$. 
Substituting into (\ref{HDE_1D}) and applying the constraint (\ref{HDE_boundary_2}) we obtain the solution:
\begin{align}
    u_0(z) &= a_0 \,, \\
    u_n(z) &= a_n e^{-(1+\i) \beta_n z } \,, \\    
 u_{-n}(z) &= a_{-n} e^{(\i-1) \beta_n z} \,, 
\end{align}
where $\beta_n \equiv \sqrt{|n|\omega \over 2\alpha}$. 
Notice that $\beta_1$ is the reciprocal of the thermal skin depth $L$ introduced in Section~\ref{sec:coordinates}, hence the $\sqrt{2}$ factor.
We determine constants $a_n$ from the boundary condition (\ref{HDE_boundary_1}). 
Here we encounter problems with the non-linear term $u^4$. 
A substitution of the Fourier sum for the temperature $u$ would lead to a~set of non-linear equations, which --- unsurprisingly --- does not have an analytical solution. 
Nevertheless, we can solve the problem analytically under the assumption that 
the changes of the temperature are significantly smaller than its absolute value. 
In such a case, we can linearize the fourth power $u^4 \approx u_0^4 + 4u_0^3 \sum_{n\neq 0} u_n e^{\i n\omega t}$.
The boundary condition (\ref{HDE_boundary_1}) then composes a~set of \emph{linear} and separated equations for the coefficients $a_n$. 
We immediately see the solution for the constant term:
\begin{equation}
    u_0 =  \sqrt[4]{\mathcal{F}_0 \over \epsilon\sigma} \,.
\end{equation}
It is convenient to introduce auxiliary parameters that help us to get rid of complex numbers in the solution. 
First, we present the thermal parameter $\Theta_n$ of the $n$-th mode:
\begin{equation}
    \Theta_n \equiv { K \beta_n \over 4 \epsilon \sigma u_0^3} \,.
\end{equation}
Once again, $\Theta_1$ corresponds to the thermal parameter $\Theta$ defined in Section~\ref{sec:coordinates}.
Second, let the phase shift $\varphi_n$ of the $n$-th mode be:
\begin{equation}
    \tan \varphi_n = - \left( {\Theta_n \over \Theta_n +1} \right) \operatorname{sgn} n\,.
\end{equation}
Since the solution is periodic, we can choose the initial time arbitrarily.
Therefore, we choose the insolation function in the form of cosine series:
$\mathcal{F}(t) = (1-A) \Phi \,\Xi(\cos \omega t)$, where $\Xi(x) = x$ for $x\geq 0$, $\Xi(x) = 0$ for $x<0$. First four Fourier modes of this function are: 
\begin{equation}
    \mathcal{F}(t) \approx (1-A) \Phi \left({1 \over \pi }+ {1 \over 2} \cos \omega t + {2 \over {3\pi}} \cos 2\omega  t - {2 \over {15 \pi}} \cos 4\omega t\right)\!.
\end{equation}

Now, we can write the solution $u$ of the problem in a simple form:
\begin{equation}    
u(z,t) = \sqrt[4]{\mathcal{F}_0 \over \epsilon \sigma} + \sum_{n=1}^\infty {\mathcal{F}_n \over 2\epsilon\sigma u_0^3} { e^{-\beta_n z} \cos \left(n\omega t - \beta_n z + \varphi_n\right)
 \over \sqrt{2\Theta_n^2 + 2\Theta_n + 1}} \,.\label{eq:u_theory}
\end{equation}
We see that for each cosine term in the series of the insolation function $\mathcal{F}$ 
there exists a~corresponding cosine term in the solution $u$, but with some phase shift. 
For the surface temperature in particular, the offset of the $n$-th Fourier mode is equal to $\varphi_n$, defined above.


\section[]{Variants of the insolation function and a~comparison with the existing one-dimensional model}
\label{sec:comparison}

\begin{figure}
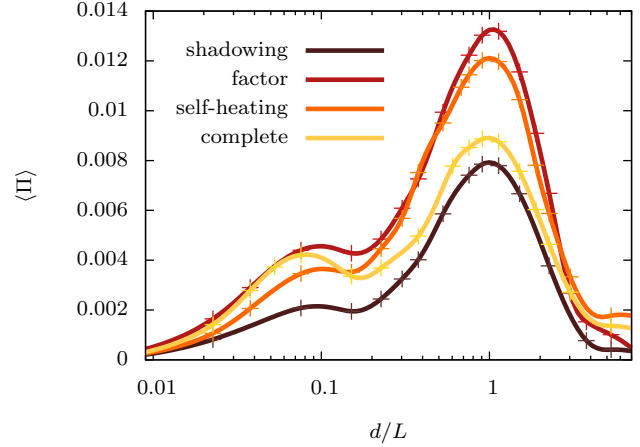

\input wall_comp
\caption{The mean dimensionless pressure~$\langle\Pi\rangle$ as a
function of the dimensionless width~$d/L$ for 
four different variants of our model for an idealized boulder:
i)~shadowing only,
ii)~factor,
iii)~self-heating, and
iv)~a complete model.
See text for a detailed explanation.}
\label{img:wall}
\end{figure}

The influence of topographic features on rotational dynamics of an asteroid has been studied by \cite{Golubov_2012}. 
Their model uses a one-dimensional approximation, 
which implicitly corresponds to a~``wall'' of an infinite height. 
Such an infinite domain cannot be used in our numerical model; 
therefore, we used the wall the height of which is about three times the width.  
The vertical heat diffusion along the wall is therefore significantly reduced (in comparison to the boulder studied in Section~\ref{sec:Pi}).
Furthermore, we employed four variants of our model, called i)~shadowing, ii)~factor, iii)~self-heating and iv)~complete. 
They can be understood as sorted by a~degree of their completeness. 

The first variant is called the ``shadowing'' model. 
It takes into account shadows casted by the boulder only.  
The self-heating effect is ignored, the influence of absorption on the~direction of the~recoil force is neglected as well. 
The insolation function is simply:
\begin{equation}
\mathcal{F} = (1-A)\Phi \mu \vec s \cdot \vec n \,.
\end{equation}

The second model, which we call the ``factor'' model, accounts for the self-heating by including simply the factor~2 to the~solar flux. 
The corresponding insolation function is therefore:
\begin{equation}
\mathcal{F} = 2(1-A)\Phi \mu \vec s \cdot \vec n \,.
\label{E_gk}
\end{equation}
This model has been used by \cite{Golubov_2012}.

The third ``self-heating'' model computes the self-heating contribution directly by a~calculation of the~thermal emission and the scattered flux as described in Section~\ref{sec:HDE}:
\begin{equation}
\mathcal{F} = (1-A)\Phi \mu \vec s \cdot \vec n + \mathcal{F}_{\rm th} + \mathcal{F}_{\rm sc} \,,
\label{E_complete}
\end{equation}
where $\mathcal{F}_{\rm th}$, $\mathcal{F}_{\rm sc}$ are the fluxes defined by Eqs. (\ref{E_rad}), (\ref{E_sc}).

We called the fourth model ``complete'', as it contains the direct computation of the self-heating, 
but also the influence of the absorption on the~direction and magnitude of the~recoil force. 
We assume that the thermally emitted radiation falling on the~surface is absorbed and does not contribute to the torque.
The recoil force is then given by:
\begin{equation}
{\boldd \vec f \over \boldd S} = -{\epsilon \sigma u^4 \over c} \left( {2 \over 3} \vec n - 
\int_{\Gamma_1} { \cos \alpha \cos \alpha' \over \pi (\vec r - \vec r')^2 }{\vec r'- \vec r \over \|\vec r' - \vec r \| } \nu \boldd \Gamma'   \right) 
\end{equation}

The mean dimensionless pressure $\langle \Pi \rangle$ (see Section~\ref{sec:pressure}) 
as a~function of the dimensionless width $d / L$ of the wall is shown in Figure \ref{img:wall}.
We notice certain common properties of the functions.  
All of them have a~global maximum at $d \sim L$, 
but there is also either a~secondary local maximum or an inflection at $d \sim 0.1\,L$. 
These properties are evident for an irregular boulder as well, as we pointed out in Section~\ref{sec:Pi}.

The maximal value of the ``factor'' curve is $\sim 0.014$, which is a~result consistent with the findings of \cite{Golubov_2012}. 
We also see that this model is a~remarkably viable approximation of the ``self-heating'' model. 
We thus confirm that the multiplication of the solar flux by the factor~2 leads to a~similar outcome 
as the inclusion of the self-heating effect (which is much more computationally demanding).

Let us compare the ``shadowing'' and ``self-heating'' models. We see a~notable property of the local YORP effect: 
the self-heating effect causes an increase of the mean pressure by $\sim 50\,\%$ in our model. 
The YORP effect induced by boulders therefore qualitatively differs from the global-shape YORP effect, 
where an inclusion of the self-heating effect leads to a~decrease of the torque magnitude for most cases \citep{Rozitis_2013}.

Finally, we compare the ``shadowing'' and ``complete'' model. 
We observe that the ``complete'' model is well approximated by the ``shadowing'' model in the vicinity of the global maximum, 
although it differs significantly for smaller boulders.
Nevertheless, this result allowed us to proceed with the simple ``shadowing'' model.
Regarding the conclusions of this paper, a~computation with the complete model would lead to very similar results.


\section[]{The numerical uncertainty of the finite element method}
\label{sec:uncertainty}

\begin{figure}
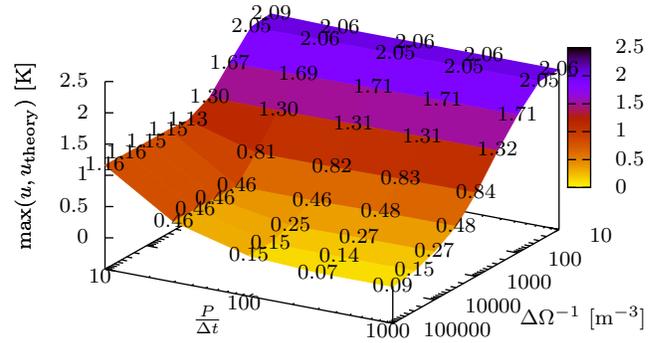

    \input convmax
    \caption{The maximal absolute difference of the numerical and analytical
solution, plotted as a~function of the discretization parameters ---
the maximum volume~$\Delta\Omega$ of tetrahedra and the time step~$\Delta t$.
The limit $\delta$ of the iterative procedure is constant, $\delta=10^{-4}$.}
    \label{img:convmax}
\end{figure}
The discretization of both space and time inevitably introduces a~numerical uncertainty to the solution. 
In order to estimate this uncertainty and find its upper bound, 
we utilize the linearized analytical solution of the HDE, 
derived in Appendix~\ref{sec:analytic}.
We select a~simple block as a~domain for our numerical model, 
i.e. we use a flat surface without any boulders or other surface features.

We are aware that the analytical solution is valid  
only if the amplitude of temperature is significantly smaller than its mean value.
That essentially corresponds to high values of the thermal conductivity $K$ (more precisely, of the thermal parameter $\Theta$).
Namely, we select such parameters that the average temperature is $\sim 180\,\rm K$ 
and the changes of temperature reach up to $\sim 6 \,\rm K$. 
In such a case, the linearized analytical solution is a~sufficiently good approximation.

The numerical uncertainty is influenced by the selection of three parameters: 
\begin{enumerate}
    \item The spatial discretization parameter $\Delta \Omega$,  
which corresponds to the upper bound for the tetrahedra volume in the three-dimensional mesh.
We use this parameter during the generation of the mesh by the TetGen code \citep{TetGen}.
\item The time step $\Delta t$. If $P$ denotes the rotational period, then $P / \Delta t$ time steps are evaluated during one period. 
\item The limit $\delta$ of the iterative process. 
    Since the problem is non-linear, we use an iterative method 
    and solve a~given sequence of linear problems in each time step. 
    We stop the process when the relative difference of two consecutive solutions is less than $\delta$.
\end{enumerate}

We measure the uncertainty by the metric:
\begin{equation}
    \max(u, u_{\rm theory}) = \max_j  \left| u^j - u^j_{\rm theory} \right| \,,
\end{equation}
where $u^j$ denotes the temperature in the $j$-th time step.
Figure \ref{img:convmax} shows the dependence of this metric
on the tetrahedra volume~$\Delta \Omega$ 
and the time step~$\Delta t$; the limit $\delta$ is a~constant here.
We checked not only the maximum difference of~$u-u_{\rm theory}$
but also its mean dispersion, nonetheless, the results were
very similar.
We see that the uncertainty indeed decreases with a refinement of the~discretization. 
We use $P/\Delta t = 1000$ and $\Delta\Omega^{-1} = 100000$ in our analyses.

For high values of the thermal conductivity~$K$,
the changes of temperature are small and the iterative procedure converges quickly --- 
the limit $\delta=10^{-4}$ was reached after only two or three iterations. 
For lower values of~$K$, the procedure with the same~$\delta$ took about 10~iterations. 
As we foreshadowed in Section~\ref{sec:HDE}, 
the procedure did not converge for the lowest values of considered range of~$K$
and we had to solve the problem using the relaxation method. 

Moreover, we have two possibilities of \emph{a posteriori} uncertainty estimates. 
First, we can compare our results to \cite{Golubov_2012} (see Appendix~\ref{sec:comparison}).
Although our and their approaches are substantially different, both results are indeed comparable. 
Second, the mean torque induced by a boulder must vanish in the limit of zero conductivity,
$K \rightarrow 0$, as we mentioned in Section~\ref{sec:orientations}. 
This is especially important as the low-$K$ case cannot be tested with the analytical solution~(\ref{eq:u_theory}).
Luckily, low~$K$ effectively corresponds to larger boulders ($\ell/L$)
which do \emph{not} contribute much to the total torque,
so we consider this case as being of lesser importance.


\bsp

\label{lastpage}

\end{document}